\documentclass[conference]{IEEEtran}
\IEEEoverridecommandlockouts
\usepackage{cite}
\usepackage{amsmath,amssymb,amsfonts}
\usepackage{algorithmic}
\usepackage{graphicx}
\usepackage{textcomp}
\usepackage{xcolor}
\usepackage{multirow}
\usepackage{comment}
\usepackage{multicol}
\usepackage{multirow}
\usepackage{xcolor}
\usepackage{arydshln}
\usepackage[top=0.75in, bottom=1.1in, left=.635in, right=.64in]{geometry}

\def\BibTeX{{\rm B\kern-.05em{\sc i\kern-.025em b}\kern-.08em
    T\kern-.1667em\lower.7ex\hbox{E}\kern-.125emX}}
\begin{document}

\title{Generalizing UxV Network Control Optimization with Disruption Tolerant Networking}



\author{\IEEEauthorblockN{1\textsuperscript{st} Quyen Dang}
\IEEEauthorblockA{\textit{Computer Science Department} \\
\textit{Naval Postgraduate School}\\
Monterey, California, USA \\
quyen.dang@nps.edu}
\and
\IEEEauthorblockN{2\textsuperscript{nd} Geoffrey Xie}
\IEEEauthorblockA{\textit{Computer Science Department} \\
\textit{Naval Postgraduate School}\\
Monterey, California, USA \\
xie@nps.edu}
}

\maketitle

\begin{abstract}
Military and disaster relief operations increasingly rely on unmanned vehicles (UxVs). It is important to develop a network control system (NCS) that can continuously coordinate and optimize the movement of UxVs based on mission objectives.  However, prior research on NCS aims to always maintain a connected network topology, which limits the utility of the resulting systems. 

In this paper, we present an approach to systematically increase the topology flexibility for an NCS  by leveraging the well-studied concept of disruption-tolerant networking (DTN). We design a DTN-compatible communication utility model that, while allowing some nodes to temporarily disconnect from others, provides for a fine-grain specification of the minimum communication frequency and the maximum hops permitted for message delivery between each pair of nodes. As such, the model supports what-if analyses prior to a mission to determine the best communication parameters to use for a given set of UxVs. Furthermore, we incorporate our communication model into an existing NCS and evaluate its performance in a simulated scenario involving the use of five UxVs in search of an enemy ship. The results show that our model not only enables the NCS to find the enemy ship faster, but also facilitates new capabilities, such as dividing the UxVs into multiple teams responsible for different search areas. 

\end{abstract}


\section{Introduction}

Autonomous systems consisting of UxVs have become critical assets in combat and disastrous relief missions. They enable access to contested environments, provide real-time situational awareness, and aid in search and rescue tasks. However, their deployment may face significant communication challenges, including physical obstructions and range limitations, necessitating resilient and adaptive communication strategies to ensure mission effectiveness. 

One such strategy is to develop a network control system (NCS) that can continuously coordinate and optimize the movement of UxVs based on mission objectives.  Notably, prior research on NCS has developed an optimization framework that uses sub-modularity utility functions to quantify the effectiveness of sensing a set of targets and, at the same time, maintaining data communications among the UxVs.~\cite{Wachlin18, Lowry20, Horner_BION}.  The framework has produced a heuristic solution that can search for the best future UxV locations in polynomial time.  

However, we observe that the prior framework aims to maintain a connected network topology \emph{at all times}, which may limit the performance of the resulting systems.  For example, in wide-area search operations, assets may be constrained from expanding coverage if they must remain within communication range, reducing searching effectiveness. In this paper,  we explore an approach to systematically increasing the topology flexibility for an NCS,  by leveraging the well-studied concept of disruption-tolerant networking (DTN). 


More specifically, we present a DTN-compatible communication utility model that precisely specifies the minimum communication frequency and maximum allowable communication hops between each pair of two nodes. These fine-grain parameters permit some nodes to temporarily disconnect from others, while ensuring periodic communication to maintain situational awareness.  Furthermore, the model supports what-if analyses, prior to a mission, to determine the best communication parameters to use for a given set of UxVs. 

We have incorporated our communication utility model into an existing NCS~\cite{Horner_BION,Lowry20} and evaluated its performance in a simulated scenario involving the use of five UxVs in search of an enemy ship. The results show that our model not only enables the NCS to find the enemy ship faster than prior work, but also facilitates new capabilities, such as dividing the UxVs into two teams responsible for different search areas. 



The remainder of the paper is organized as follows. Section II provides a tutorial of the prior NCS framework that has motivated this work, and a brief review of DTN. Section III presents the proposed DTN communication utility function. Then, the DTN communication utility function is evaluated through a simulation study, as detailed in Section IV. Finally, Section V concludes the paper with a summary and potential directions for future work.

\vspace{-5pt}
\section{Related Work}
Our review of related work consists of two parts. First, we provide a detailed tutorial of multi-objective NCS optimization utilizing submodularity utility functions. The information is critical for understanding the evaluation part of our work (Section~\ref{sec:evaluate}), which leverages the same optimization framework. 
Second, we briefly summarize core concepts and sample applications of DTN.

\nocite{lee2020}

\subsection{NCS Optimization with Submodular Utility Functions}
\label{sec:optimzation}

Horner~{\em et al.}\cite{Horner23,Wachlin18,Lowry20} investigated the problem of optimizing the placement of UxV assets in an NCS over time to meet both the communication and sensing requirements. They discretized the geographic area as a 2D or 3D  grid map, (e.g., a $100\times100$ grid with a maximum search space of 10,000 points). They also discretized the time of operation into control cycles (e.g., 30-minute intervals): New UxV positions were determined at the beginning of each control cycle. In the following, we will present the main parts of this optimization framework, which we will refer to simply as ``the framework'' in the rest of the section. 

{\bf Formulation of Optimization Problem:} An NCS generally consists of both UxVs and manned vehicles. The framework abstracts all vehicles, manned or unmanned, as nodes of the same type that form a graph. One or more nodes will move according to preset requirements (e.g., a command vehicle loitering over an area or a convoy of manned vehicles following a path toward a target), while the other nodes need to react to the preset movement(s) and adjust their positions every control cycle. At the core of the NCS optimization problem is how to reposition the ``reactive'' nodes, per control cycle, in order to optimize one or more performance objectives. 

The framework primarily considers two performance objectives: (i) \emph{sensing}, i.e., the ability to monitor a set of stationery or moving targets; and (ii) \emph{communications}, i.e., the ability for the nodes to communicate with each other. It models each of the objectives with a utility function over an NCS topology. To reduce the complexity of the optimization problem, it combines the two objectives into one joint optimization criterion by taking a weighted sum of the sensing and communications utilities, as follows. Let $f_s()$ and $f_c()$ denote the sensing and communications utility functions, respectively. The joint utility function for a given NCS topology $G$ is modeled as:
\begin{equation}
    J(G) = \alpha_s\times f_s(G) + \alpha_c\times f_c(G)    
    \label{eq:joint-utility}
\end{equation}
where $\alpha_s$ and $\alpha_c$ are positive weights such that $\alpha_s + \alpha_c = 1$. As such, the framework is about finding the next optimizing topology, denoted by $G^*$ and defined by the equation below, in each control cycle.
\begin{equation}
  G^* = \operatorname*{argmax}_{G\in pos(V)} J(G)
\end{equation}
where $V$ is the set of all reactive nodes, and $pos(V)$ denotes a function that enumerates \emph{all feasible NCS topologies} from repositioning each node in $V$ once. 

Searching for the best positions of multiple nodes simultaneously to maximize the joint utility $J$ would introduce a combinatorial problem with an exponential growth of computational complexity as the size of $V$ increases. The framework mitigates this problem using a well-known greedy heuristic~\cite{nemhauser78,krause14} that can obtain a near-optimal $J$ value by determining the best node positions sequentially, i.e., \emph{one node at a time}.

The heuristic requires that the sensing and communications utility functions are submodular.~\cite{krause14} 
In particular, the framework utilizes a subclass of submodular functions, called monotone submodular functions. Formally, an NCS utility function $f:2^{V}\mapsto\mathbb{R}$ with respect to a set of UxV nodes $V$ is monotone submodular if and only if $\forall A\subseteq B \subseteq V$, the following holds:
\vspace{-3pt}
\begin{equation}
    \vspace{-3pt}
     f(A) \leq f(B)    
    \label{eq:monotone}
\end{equation}
Note that when both the sensing and communications utility functions are monotone submodular, the joint utility function $J()$ is also monotone submodular. Intuitively, with a monotone $J()$ function, repositioning an additional (new) node will \emph{never} cause the joint utility to decrease. Therefore, in each control cycle, the framework is able to achieve a near optimal $J$ value by iterating through the nodes \emph{in the order of the amount of maximum utility increase} they can achieve by moving to a new feasible location.~\cite{Wachlin18,Lowry20} 

\vspace{3pt}

{\bf Sensing Utility Function:} The framework models the sensing utility of a UxV at a potential location based on how effectively the UxV is able to monitor a set of predetermined targets of interest.~\cite{Wachlin18,Lowry20} Consider a UxV node $i$ and a target $j$. Let $r$ be their distance in the 2D grid space if the UxV is moved to that location. The framework calculates the new sensing utility, denoted by $\theta_{i, j}(r)$, using the equation below:
{\small
\begin{equation}
    \theta_{i, j}(r)= 
    \begin{cases}
        w_j, & 0 \leq r \leq s_{\text{suf}} \\
        \displaystyle\frac{w_j}{2}\left(1 + \cos\left(\pi \frac{r - s_{suf}}{s_{max} - s_{suf}} \right)\right), & s_{suf} < r \leq s_{max} \\
        0, & r > s_{\text{max}}
    \end{cases}
    \label{eq:sensing}
\end{equation}
}

\noindent where $w_j$ is a positive constant defining the maximum sensing value (i.e., relative importance) of the target, $s_{suf}$ indicates the minimum $r$ that is for sufficient to achieve the full utility of $w_j$, and $s_{max}$ the maximum $r$ beyond which there is no sensing utility. Note that $\theta_{i, j}(r)$ gradually degrades from $w_j$ to 0 when $r$ falls between $s_{suf}$ and $s_{max}$. 

The framework then derives the overall sensing utility of the UxV location by summing up the per target sensing utilities, as stated in equation~(\ref{eq:sensing}),  over all targets.~\cite{Wachlin18}

\vspace{3pt}

{\bf Communications Utility Function:} The framework models the change of communication utility after repositioning a UxV at a new location using the concept of graph resistance.~\cite{Wachlin18}  Consider a UxV node $i$ and another node $j$. When node $i$ moves, the distance between the two nodes, denoted by $d$, will change. The framework calculates the edge resistance between $i$ and $j$, denoted by  $e_{i, j}(d)$, using the equation below:
{\small
\begin{equation}
    e_{i, j}(d)= 
    \begin{cases}
    1.0 - 0.9e^{-\tau(d/c_{max})},&    {0 < d\leq c_{max}}\\
    \infty, & {d > c_{max}}\\
    \end{cases}
    \label{eq:edge_weight}
\end{equation}
}

\noindent where $c_{max}$ is the maximum feasible communication range and $\tau$ a positive constant parameter. Note that the edge resistance increases inverse exponentially as long as $d$ does not exceed $c_{max}$. After $d$ exceeds $c_{max}$, the resistance is set to $\infty$ to indicate that communication is infeasible.

The framework then considers the new NCS topology after repositioning node $i$ as a weighted graph with the weight of each edge connecting $i$ updated with equation~(\ref{eq:edge_weight}), and derives the overall graph resistance, denoted by $R$, using a well-known equation~\cite{ellens2013} based on the eigenvalues of the graph's Laplacian matrix, as follows. 
\vspace{-4pt}
\begin{equation}
\vspace{-3pt}
     R = N\sum_{k=2}^{N}\frac{1}{\lambda_k}
     \label{eq:R}
\end{equation}
where $N$ is the matrix's dimension, and $\lambda_k$ the matrix's $k$th eigenvalue, $k=2,...,N$.  

Finally, the framework normalizes the set of $R$ values corresponding to all feasible locations for repositioning node $i$ in this cycle so that $R$ is upper bounded by 1, and uses $1 - Norm(R)$ as the the communication utility for repositioning node $i$ to the new location.

Note that an implicit assumption for using equation~(\ref{eq:R}) is that the NCS is a connected graph. When the graph after the repositioning is disconnected, the framework sets the resulting communication utility to zero. This severe penality for disconnections, even when they only last for a small number of control cycles, limits the utility of the framework.  

\subsection{DTN Concept and Applications}

DTN\footnote{DTN may also refer to delay tolerant networking, which has been explored in the context of inter-planetary communication networks and underwater acoustic networks.} is a communication paradigm developed to support reliable message delivery in environments with intermittent connectivity. Unlike conventional networking protocols, DTN employs a \textit{store-and-forward} mechanism in which messages -- encapsulated as bundles -- are stored at intermediate nodes and forwarded opportunistically when connections become available~\cite{Rohrer_Xie,Warthman_2015}. This design was originally conceived by NASA to address the limitations of traditional protocols in deep-space communication, where planetary movement and extreme distances introduces substantial delay and disruption, rendering standard TCP/IP protocol ineffective~\cite{Warthman_2015, McMahon_Farrell}.

Beyond space applications, DTN has been adapted for terrestrial use, particularly in scenarios where stable network infrastructure is unavailable. One example is the CAMON system developed by researchers at the Technical University of Darmstadt (Germany), which integrates DTN with UAVs to monitor disaster areas. In these environments, UAVs collect and share network information using DTN protocols, improving both detection speed and data integrity despite limited connectivity~\cite{Zobel_Kundel_Steinmetz}. Another study from the University of Electro-Communications (Japan) proposed a DTN routing protocol that uses UAVs as mobile relays. By leveraging historical encounter data to optimize relay selection, the protocol improved delivery rates and reduced latency in simulated urban settings, demonstrating the promise of DTN in delay-prone, infrastructure-limited environment~\cite{Du_Wu_Tsutomu}.

\section{Design of DTN Communication Utility}

This section presents the design of a new DTN motivated communication utility function for NCS optimization, which we call the $Net()$ model. First, we will describe the key parameters defined by the model and the associated link prioritization requirement, leading to the derivation of a new communication utility function. Then, we will discuss a potential teaming extension that supports the UxVs to perform multiple tasks in parallel. 

\nocite{Dang24}

\subsection{The $Net()$ Model}

{\bf Communication parameters}: As discussed earlier, nodes in a DTN network can exchange messages across one or more hops of intermittent connections. We model such flexibility on a per node pair basis. More specifically, for each node pair $(i, j)$ in an NCS, we formally define the pair's communication requirement as $Net(i,j)$, which consists of two parameters, as follows: 
\vspace{-2pt}
\begin{equation}
   Net(i,j) = \{c(i,j), h(i,j)\}    \label{eqn:net-1}
    \vspace{-2pt}
\end{equation}
where $c(i,j)$ denotes \emph{the maximum number of control cycles} during which nodes $i$ and $j$ are unable to communicate with each other, and $h(i,j)$ the \emph{maximum number of hops} allowed for data transfer between nodes $i$ and $j$ when they can communicate.

For example, when $c(i,j) = 3$ and $h(i,j) = 4$, nodes $i$ and $j$ are allowed to be disconnected for up to three control cycles and it should take no more than four hops (i.e., via up to three relay nodes) to deliver messages between them. Note that $h(i,j)$ impacts the permitted message delivery latency and should be chosen accordingly.   

In particular, we have chosen to define $c()$ and $h()$ parameters on a fine-grain granularity of node pairs to support ``what-if'' analyses that can determine, prior to a mission, the best communication parameters to use for a given set of UxVs. As will be shown in Section~\ref{sec:eval_base}, such analyses can be very effective in improving the NCS performance. 

{\bf Link prioritization}: Based on the $Net()$ model, we can define the communication utility of a candidate NCS topology in control cycle $k$, denoted by $G_k$. First, we define a Boolean function for each pair of nodes $i$ and $j$ to indicate whether $G_k$ conform to the communication requirements for the two nodes as specified in $Net(i,j)$:
\begin{equation}
    m(i,j,G_k) = 1, \mbox{iff $G_k$ conforms to $Net(i,j)$}
    \label{eq:m}
\end{equation}
\noindent Intuitively, to determine $m(i,j,G_k)$ correctly requires tracking (i) the previous control cycles when $i$ and $j$ were disconnected and (ii) the shortest path between $i$ and $j$ in each control cycle when they are connected. For brevity, we omit the details. 

Due to sporadic disconnections inherent in DTNs, a prioritized communication structure is necessary to guide node decisions when multiple communication links yield the same utility values. This approach ensures that time-sensitive or mission-critical information, such as situational updates or sensor data relevant to operational decision making, is prioritized over routine status exchanges between systems. 

To prioritize communication links, we assign weights to node pairs based on their required communication frequency. Since lower $c(i, j)$ values indicate higher communication priority, we use the inverse of $c(i, j)$ as weights to emphasize more critical links. These weights are used as scale factors for the pairwise utility values (i.e., $m(i,j,G_k)$). And the sum of  scaled pairwise utility values is normalized by the total weights to ensure that the communication utility value has a range between 0 and 1, as shown below.
    
\vspace{-4pt}
\begin{equation}\label{eq:Quyen_f_r}
   f_{c}(G_k)= \frac{\sum_{i < j \leq N} (\frac{1}{c(i,j)} \times m(i, j, G_k))}{\sum_{i< j \leq N} \frac{1}{c(i,j)}}
\end{equation}

It is straightforward to show that the above function is monotone submodular, and thus, is compatible with the optimization framework presented in Section~\ref{sec:optimzation}. 

\subsection{Teaming Extension}
\label{sec:teaming}

In some reconnaissance scenarios, multiple targets -- such as enemy locations or valuable assets -- may be dispersed across a wide area. It is desirable to divide UxVs into teams, each of which focuses on a different target, to improve the utilization of these platforms. Thus, we investigate a teaming extension, where UxVs are explicitly organized into separate teams for parallel tasking.

To support teaming, the $Net()$ parameters are adjusted so that intra-team node pairs have lower $c()$ values than inter-team node pairs. This indicates a requirement for more frequent communication between members of the same team and leads to natural partitioning, with dense links forming within teams and sparse links across teams.  It should be noted that cross-team links are part of the overall NCS topology and can be used for relaying messages from all nodes. 

Additionally, the node and target specific sensing utility calculated with equation~(\ref{eq:sensing}) is scaled with a priority weight. In scenario with two targets, a higher weight value (e.g., 0.8 vs.\ 0.2) is assigned when the target falls under the node's team responsibility. Such scaling is designed to direct each team formation to move toward its designated team target.





\section{Evaluation}\label{sec:evaluate}

\vspace{-2pt}
\begin{figure}[htb]
    \vspace{-5pt}
    \centering
    \includegraphics[width=0.7\linewidth]{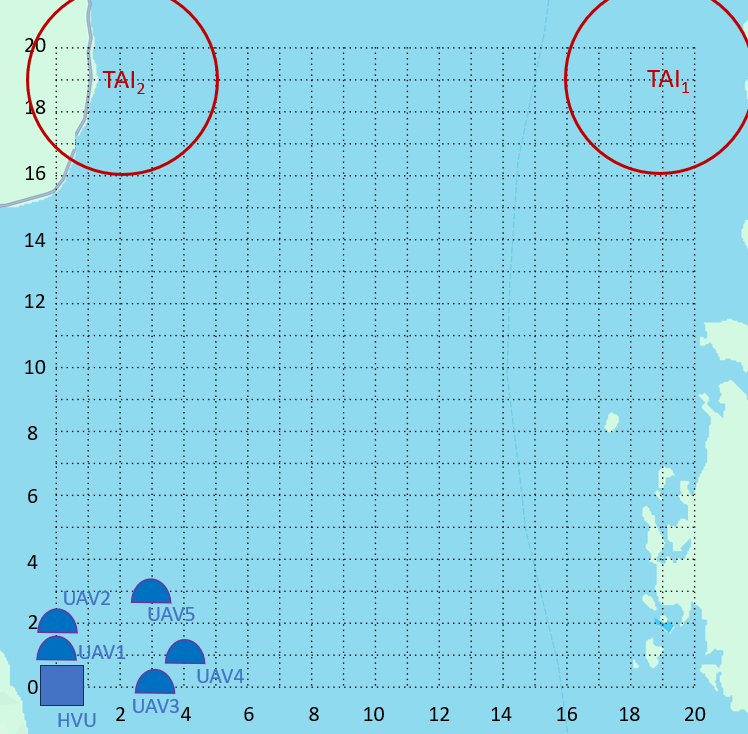}
    \caption{Simulation scenario: Operation space discretized into 20x20 grid.}
    \label{fig:map}
    \vspace{-10pt}
\end{figure}
Our simulation based evaluation is divided into two parts. The first evaluates the utility of the $Net()$ model by conducting a ``what-if'' analysis with three sample $Net()$ configurations, while the second validates the teaming extension's more efficient use of UxVs.

We simulate a maritime search scenario as illustrated in Fig.~\ref{fig:map}. In this setup, Blue and Red forces are initially positioned beyond each other's detection range. The Blue force's objective is to locate a Red target while keeping the High Value Unit (HVU) ship undetected. Five UAVs are deployed to extend the search toward Target Areas of Interest (TAIs) identified from the last known intelligence reports. The operation space is represented as a $20\times20$ grid, where each intersection corresponds to a potential node position.
For simplicity, all grid points are assumed to be navigable water. The simulation code base, originally developed in MATLAB~\cite{Wachlin18,Lowry20}, was converted to Python and modified to incorporate the $Net()$ model, resulting in 1219 lines of Python code.

We assume the UAVs to have the same sensing range ($s_{max}$), communication range ($c_{max}$), and maneuvering range ($m_{max}$), as provided in Table~\ref{tab:param}. The parameters are measured by grid steps. For example, if each grid step is 500
meters, then $s_{max}$ = 4 × 500 = 2000 meters.

\begin{table}[htb]
\vspace{-2pt}
    \centering
    \textbf{UAV capability parameters}\\
    \vspace{2pt}
    \begin{tabular}{|c|c|c|}
         \hline
         $s_{max}$ & $c_{max}$ & $m_{max}$ \\
         \hline
         4 & 5 & 4 \\
         \hline
    \end{tabular}
    \vspace{3pt}
    \\
    \caption{}
    \label{tab:param}
\end{table}

\vspace{-20pt}

\subsection{Evaluation of Base $Net()$ Model}\label{sec:eval_base}

Table~\ref{tab:Nets} displays the three $Net()$ configurations used in this part of the evaluation. 
Given that the operation space is small relative to the communication and maneuvering ranges of the UAVs, we relax the hop limit constraints for most of the node pairs by setting their $h()$ parameters to 10. 


\begin{table}[htb]
\vspace{-5pt}
    \centering
    \begin{tabular}{rc}
        \multicolumn{2}{c}{\textbf{Net-1 Configuration}}\\
        &  \text{HVU} ~~~ \text{UAV1} ~~ \text{UAV2} ~~~  \text{UAV3} ~~ \text{UAV4} ~~ \text{UAV5}\\ 
        \text{HVU} & \multirow{6}{*}{$\left[
        \begin{array}{cccccc}
            -      & (1,1)  & (2,10) & (2,10) & (3,10) & (3,10) \\
            (1,1)  & -      & (1,1)  & (2,10) & (3,10) & (3,10)\\
            (2,10) & (1,1)  & -      & (1,1) & (2,10) & (3,10)\\
            (2,10) & (2,10) & (1,1) & -      & (3,10) & (3,10)\\
            (3,10) & (3,10) & (2,10) & (3,10) & -      & (1,1)\\
            (3,10) & (3,10) & (3,10) & (3,10) & (1,1) & - \\
        \end{array}
        \right]$}\\
        \text{UAV1} & \\
        \text{UAV2} & \\
        \text{UAV3} & \\
        \text{UAV4} & \\
        \text{UAV5} & \\
        \\
        \multicolumn{2}{c}{\textbf{Net-2 Configuration}}\\
        &  \text{HVU} ~~~ \text{UAV1} ~~ \text{UAV2} ~~~  \text{UAV3} ~~ \text{UAV4} ~~ \text{UAV5}\\ 
        \text{HVU} & \multirow{6}{*}{$\left[
        \begin{array}{cccccc}
            -      & (1,1)  & (2,10) & (2,10) & (3,10) & \textbf{(4, 10)} \\
            (1,1)  & -      & (1,1)  & (2,10) & (3,10) & \textbf{(4, 10)}\\
            (2,10) & (1,1)  & -      & (1,1) & \textbf{(3, 10)} & \textbf{(4, 10)}\\
            (2,10) & (2,10) & (1,1) & -      & (3,10) & \textbf{(4, 10)}\\
            (3,10) & (3,10) & \textbf{(3, 10)} & (3,10) & -      & (1,1)\\
            (3,10) & (3,10) & (3,10) & (3,10) & (1,1) & - \\
        \end{array}
        \right]$}\\
        \text{UAV1} & \\
        \text{UAV2} & \\
        \text{UAV3} & \\
        \text{UAV4} & \\
        \text{UAV5} & \\
        \\ 
        \multicolumn{2}{c}{\textbf{Net-3 Configuration}}\\
        &  \text{HVU} ~~~ \text{UAV1} ~~ \text{UAV2} ~~~  \text{UAV3} ~~ \text{UAV4} ~~ \text{UAV5}\\ 
        \text{HVU} & \multirow{6}{*}{$\left[
        \begin{array}{cccccc}
            -      & (1,1)  & (2,10) & \textbf{(3, 10)} & \textbf{(4, 10)} & \textbf{(5, 10)}\\
            (1,1)  & -      & (1,1)  & \textbf{(3, 10)} & \textbf{(4, 10)} & \textbf{(5, 10)}\\
            (2,10) & (1,1)  & -      & \textbf{(3, 10)} & \textbf{(4, 10)} & \textbf{(5, 10)}\\
            \textbf{(3, 10)} & \textbf{(3, 10)} & \textbf{(3, 10)} & -      & \textbf{(4, 10)} & \textbf{(5, 10)}\\
            \textbf{(4, 10)} & \textbf{(4, 10)} & \textbf{(4, 10)} & \textbf{(4, 10)} & -      & \textbf{(5, 10)}\\
            \textbf{(5, 10)} & \textbf{(5, 10)} & \textbf{(5, 10)} & \textbf{(5, 10)} & \textbf{(5, 10)} & - \\
        \end{array}
        \right]$}\\
        \text{UAV1} & \\
        \text{UAV2} & \\
        \text{UAV3} & \\
        \text{UAV4} & \\
        \text{UAV5} & \\
        \\
    \end{tabular}
    \caption{}
    \label{tab:Nets}
\vspace{-13pt}
\end{table}

\begin{figure*}
    \centering
    \includegraphics[width=0.98\textwidth]{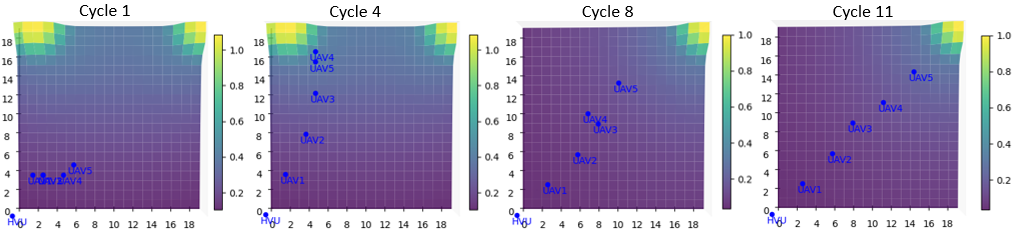}
    \caption{Net-1 Results: Simulation fails to explore TAI$_1$ due to restrictive communication constraints.}
    \label{fig:Net-1}
\end{figure*}

\begin{figure*}[htb]
\vspace{-10pt}
    \centering
    \includegraphics[width=0.98\textwidth]{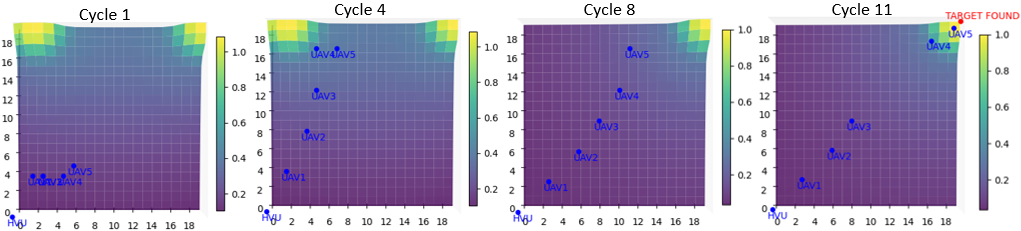}
    \caption{Net-2 Results: Relaxing the $Net()$ model enables independent node operation, allowing successful exploration of TAI$_1$ and target detection.}
    \label{fig:Net-2}
\vspace{-10pt}
\end{figure*}

The Net-1 model is designed to position nodes in a relay formation toward the TAIs, balancing search coverage and sustained network connectivity. To maintain communication feasibility across the network, the model ensures that each node retains at least one direct link to another, establishing reliable relay paths throughout the formation. In other words, this configuration \emph{does not leverage DTN}. 

Simulation results in Fig.~\ref{fig:Net-1} shows that nodes perform an iterative search, starting with TAI$_2$ where sensing utilities are higher due to proximity, and progressing to TAI$_1$ upon clearing the current area. However, as UAVs approach TAI$_1$, communication constraints inhibit their ability to disconnect from the relay structure, ultimately preventing exploration of that region. By Cycle 11, nodes converge to positions of optimal utility and remain stationary indefinitely. 

To alleviate the constraints of Net-1, the Net-2 model adjusts $c(i,j)$ for selected node pairs, as indicated in bold in Table~\ref{tab:Nets}. 
Fig.~\ref{fig:Net-2} illustrates that Net-2 successfully enables UAV4 and UAV5 to temporarily disconnect from the relay structure and operate beyond the communication range, effectively leveraging DTN to explore TAI$_1$. This results in successful target detection by Cycle 11, completing the mission.

\begin{figure*}
\vspace{4pt}
    \centering
    \includegraphics[width=0.98\textwidth]{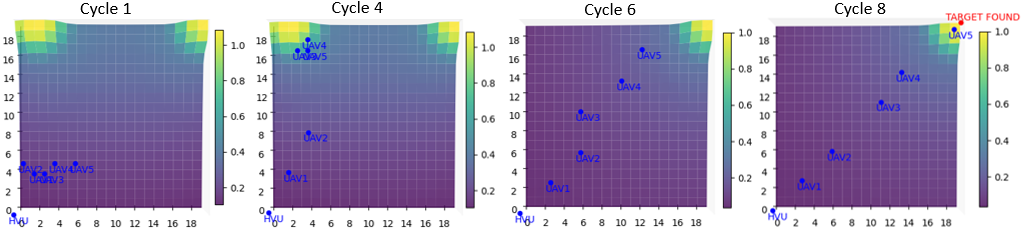}
    \caption{Net-3 Results: Modifying the $Net()$ model yields a different node formation and search pattern, enabling successful target detection within 8 cycles.}
    \label{fig:Net-3}
\vspace{-6pt}
\end{figure*}

To reduce target detection time, Net-3 adjusts several $c(i,j)$ values relative to Net-2, as bolded in Table~\ref{tab:Nets}, further relaxing communication requirements for UAV3, UAV4, and UAV5.
The result as shown in Fig.~\ref{fig:Net-3} demonstrates that UAV3, UAv4, and UAV5 disconnect from the other nodes and advance toward TAI$_1$. By Cycle 4, a distinct formation emerges, contrasting with the previous results, and ultimately enables the nodes to locate target within 8 cycles, 3 cycles faster than that of using Net-2. This highlights how different $Net()$ communication parameters yield varying behaviors: Net-2 moderately relaxes connectivity constraints, while Net-3 allows greater flexibility to further enhance search performance.

\begin{figure*}[htb]
\vspace{-2pt}
    \centering
    \includegraphics[width=0.98\textwidth]{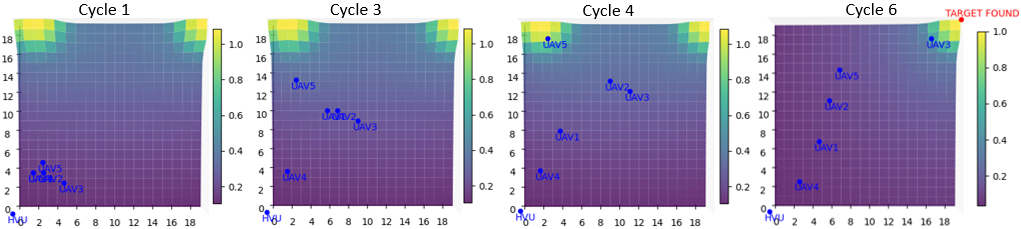}
    \caption{Net-Team Results: Team-based configuration enables parallel exploration and flexible inter-team support, achieving target detection in 6 cycles.}
    \label{fig:Net-Team}
\vspace{-10pt}
\end{figure*}

\subsection{Evaluation of Teaming Extension}

While Net-2 and Net-3 are effective, Fig.~\ref{fig:Net-2} and Fig.~\ref{fig:Net-3} show that they have limitations: Without teaming consideration, all UAVs focus on TAI$_2$ first and then TAI$_1$, causing redundant work and unnecessary energy consumption among the nodes, and potentially delaying the search process.

To evaluate the teaming extension presented in \mbox{Section} \ref{sec:teaming}, we assign the UAVs to two teams. UAV1, UAV2, and UAV3 form Team A tasked to search TAI$_1$, while UAV4 and UAV5 comprise Team B tasked to TAI$_2$ that is a shorter distance away. 
Each team uses a sensing weight value of 0.8 for their tasked TAI, and 0.2 for the other TAI. 

\vspace{-5pt}
\begin{table}[htb]
    \centering
    \resizebox{.5\textwidth}{!}{
    \begin{tabular}{rc}
        \multicolumn{2}{c}{\textbf{Net-Team Configuration}}\\
        &\text{HVU} ~~~~ \textcolor{cyan}{\text{UAV1} ~~~~ \text{UAV2} ~~~~~ \text{UAV3} ~~~~~} \textcolor{blue}{\text{UAV4} ~~~~ \text{UAV5}} \\ 
        \text{HVU} & \multirow{6}{*}{$\left[
        \begin{array}{cccc:cc}
            -      & (2,10) & (4,10) & (6,10) &(1,1)  & (6,10) \\
            (2,10) & -      & (2,10) & (6,10) &(10,10) & (10,10)\\
            (4,10) & (2,10) & -      & (4,10) & (10,10) & (10,10)\\
            (6,10) & (6,10) & (4,10) & -      & (10,10) & (10,10)\\
            \multicolumn{6}{l}{\text{-- -- -- -- -- -- -- -- -- -- -- -- -- -- -- -- -- -- -- -- -- -- -- -- -- -- -- -- -- -- --}}\\
            (1,1)  & (10,10) & (10,10) & (10,10) & -     & (4,10)\\
            (6,10) & (10,10) & (10,10) & (10,10) & (4,10) & - \\
        \end{array}
        \right]$}\\
        \textcolor{cyan}{\text{UAV1}} & \\
        \textcolor{cyan}{\text{UAV2}} & \\
        \textcolor{cyan}{\text{UAV3}} & \\
        \textcolor{black}{\text{-- -- --}} & \\
        \textcolor{blue}{\text{UAV4}} & \\
        \textcolor{blue}{\text{UAV5}} &\\
        \\
    \end{tabular}}
    \caption[Net-Team model] {UAVs assigned to Team A (cyan) and Team B (blue).}
    \label{tab:Net-Team}
\end{table}
\vspace{-15pt}

Following multiple experiments, the Net-Team configuration shown in Table~\ref{tab:Net-Team} was identified as the most effective. Node pairs within the same team are assigned lower $c(i,j)$ values (4 vs.\ 10) to prioritize intra-team connectivity and coordination 
while increasing the flexibility of spatial separation between teams. To support extended communication with the HVU, Team B designates one node as a relay, thereby improving the efficiency of intelligence data transmission.

The result presented in Fig.~\ref{fig:Net-Team} indicates a distinct separation between the two teams starting at cycle 3. Team A progresses toward TAI$_1$, while Team B simultaneously advances toward TAI$_2$. UAV4 maintains its position as a relay node for the HVU, enhancing communication range. Once UAV5 clears its current area, it proceeds toward Team A to support the search efforts. This demonstrates that the separation configuration does not inhibit nodes' ability to converge on a common objective when a single target is present, thereby optimizing resource allocation. Ultimately, this configuration enables locating the target within 6 cycles, the fastest search time achieved in our experiments.

\vspace{-5pt}
\section{Conclusion and Future Work}\label{sec:conclusion}



This study has demonstrated that DTN can systematically increase the topology flexibility of an NCS. The proposed $Net()$ model is shown to support ``what-if'' analyses that can significantly impact mission outcomes. The ability to fine-tune $Net()$ parameters can shift a scenario from failure to success and is critical for formulating strategies to further enhance UxV teaming performance. The simulation model developed offers a cost-effective and rapid means to evaluate multiple $Net()$ configurations, supporting iterative refinement and informed decision-making without expending real-world resources.



For future work, we plan to focus on automating the exploration of $Net()$ configurations to efficiently identify optimal parameter settings for a given mission scenario, thereby reducing reliance on manual selection and speeding up mission planning. Additionally, we plan to integrate energy consumption factors into the simulation framework, which would enhance realism and enable the development of energy preservation strategies. 

\begin{small}
\bibliographystyle{IEEEtran}
\bibliography{paper}
\end{small}

\end{document}